\def\Black{} 
\def\Blue{} 
\def\Brown{} 
\def\slashchar#1{\setbox0=\hbox{$#1$}           
   \dimen0=\wd0                                 
   \setbox1=\hbox{/} \dimen1=\wd1               
   \ifdim\dimen0>\dimen1                        
      \rlap{\hbox to \dimen0{\hfil/\hfil}}      
      #1                                        
   \else                                        
      \rlap{\hbox to \dimen1{\hfil$#1$\hfil}}   
      /                                         
   \fi}                                         %
\def\RPV{\slashchar{R}} 
\newcommand{\be}{\begin{equation}} 
\newcommand{\ee}{\end{equation}} 
\newcommand{\bea}{\begin{eqnarray}} 
\newcommand{\ena}{\end{eqnarray}} 
\newcommand{\eea}{\end{eqnarray}} 
 
\def\pr#1#2#3{ Phys. Rev. ${\bf{#1}}$ (#2) #3} 
 
\def\plb#1#2#3{ Phys. Lett. ${\bf{#1}}$ (#2) #3}
 
\def\np#1#2#3{ Nucl. Phys. ${\bf{#1}}$ (#2) #3} 
\def\zp#1#2#3{ Z. Phys. ${\bf{#1}}$ (#2) #3}

\documentstyle[aps,epsf,prl]{revtex}
\begin{document} 
 
\begin{titlepage} 
\null 
\vspace{0cm} 
\begin{center} 
\Large\bf  
\Brown 
Single top production at the LHC as a probe of $R$ parity  
violation 
\Black 
\end{center} 
\vspace{0.5cm} 
 
\begin{center} 
\begin{large} 
P. Chiappetta \\ 
\end{large} 
\vspace{0.3cm} 
Centre de Physique Th\'eorique, UPR7061, CNRS-Luminy, Case 907, \\  
F-13288 Marseille Cedex 9, France\\ 
\vspace{0.5cm} 
\begin{large} 
A. Deandrea \\ 
\end{large} 
\vspace{0.3cm} 
Theoretical Physics Division, CERN, CH-1211 Geneva 23, Switzerland \\ 
\vspace{0.5cm} 
\begin{large} 
E. Nagy and S. Negroni \\ 
\end{large} 
\vspace{0.3cm} 
Centre de Physique des Particules de Marseille, Universit\'e de la 
M\'editerran\'ee, \\ 
Case 907, F-13288 Marseille Cedex 9, France\\ 
\vspace{0.5cm} 
\begin{large} 
G. Polesello \\ 
\end{large} 
\vspace{0.3cm} 
INFN, Sezione di Pavia, via Bassi 6, I-27100 Pavia Italy \\ 
\vspace{0.5cm} 
\begin{large} 
J.M. Virey \\ 
\end{large} 
\vspace{0.3cm} 
Institut f\"ur Physik, Universit\"at Dortmund, D-44221 Dortmund, Germany and, \\ 
Centre de Physique Th\'eorique, UPR7061, CNRS-Luminy, Case 907, \\  
F-13288 Marseille Cedex 9, France and Universit\'e de Provence, Marseille,
France \\  \end{center} 
 
\vspace{0.5cm} 
\begin{center} 
\begin{large} 
\Brown 
{\bf Abstract}\\[0.5cm]\Black 
\end{large} 
\parbox{14cm}{We investigate the potential of the LHC  
to probe the $R$ parity violating couplings involving the  
third generation by considering single top production. This study is based  
on particle level event generation for both signal and background, interfaced to 
a simplified simulation of the ATLAS detector.} 
\end{center} 
\vspace{2cm}
\noindent
\Blue
\Black
\vfil
\noindent  
\Brown
CERN-TH/99-313 \\
CPT-99-P/3830 \\ 
CPPM-P-1999-03 \\ 
DO-TH 99/07 \\ 
ATL-COM-PHYS-99-043 \\ 
October 1999 
\Black 
\end{titlepage} 

\setcounter{page}{1}

\preprint{CERN-TH/99-313 \\
CPT-99-P/3830 \\ 
CPPM-P-1999-03 \\ 
DO-TH 99/07 \\ 
ATL-COM-PHYS-99-043}
\title{Single top production at the LHC as a probe of $R$ parity  
violation}
\author{P. Chiappetta}
\address{Centre de Physique Th\'eorique, UPR7061, CNRS-Luminy, Case 907, \\  
F-13288 Marseille Cedex 9, France}
\author{A. Deandrea}
\address{Theoretical Physics Division, CERN, CH-1211 Geneva 23, Switzerland}
\author{E. Nagy and S. Negroni}
\address{Centre de Physique des Particules de Marseille, Universit\'e de la 
M\'editerran\'ee, \\ 
Case 907, F-13288 Marseille Cedex 9, France}
\author{G. Polesello} 
\address{INFN, Sezione di Pavia, via Bassi 6, I-27100 Pavia Italy}
\author{J.M. Virey}
\address{Institut f\"ur Physik, Universit\"at Dortmund, D-44221 Dortmund, Germany and, \\ 
Centre de Physique Th\'eorique, UPR7061, CNRS-Luminy, Case 907, \\  
F-13288 Marseille Cedex 9, France and Universit\'e de Provence, Marseille,
France} 
\date{October 1999}
\maketitle
\begin{abstract}
We investigate the potential of the LHC  
to probe the $R$ parity violating couplings involving the  
third generation by considering single top production. This study is based  
on particle level event generation for both signal and background, interfaced to 
a simplified simulation of the ATLAS detector.
\end{abstract}

\section{Introduction} 
 
The conservation of the baryon $B$ and lepton $L$ number is a consequence  
of the gauge invariance and  renormalizability of the Standard Model.  
In supersymmetric extensions of the Standard Model, gauge invariance and 
renormalizability do not imply baryon and lepton number 
conservation. We shall consider in what follows the Minimal Supersymmetric 
Standard Model (MSSM) together with baryon or lepton number violating 
couplings. These Yukawa-type interactions are often referred to as  
$R$-parity violating couplings. They can mediate proton decay 
to an unacceptable level and for this reason a discrete symmetry $R$ was   
postulated \cite{fayet} that acts as $1$ on all known particles and  
as $-1$ on all the superpartners:  
\begin{equation}  
R = (-1)^{3B+L+2S} 
\end{equation} 
where $S$ is the spin of the particle. In the MSSM with a conserved $R$-parity  
the lightest supersymmetric particle (LSP) cannot disintegrate into ordinary  
particles and is therefore stable. The superpartners can be produced only 
in pairs so that one needs usually to wait for high energy colliders. 
 
In models \cite{models} not constrained by the ad-hoc imposition of $R$-parity
one can still avoid proton decay and the experimental signatures can be quite
interesting: single production of supersymmetric particles and modification
of standard decays and cross-sections due to the exchange of these
sparticles, which could be observed at lower energies compared to the
$R$-parity conserving model. In the following we shall investigate top quark
production taking into account $R$-parity violating effects. 
The top quark being heavy with a mass close to the electroweak symmetry  
breaking scale, it is believed to be more sensitive to new physics than other  
quarks. The mechanism we plan to study is single top quark production at LHC, 
which is complementary to top quark  
pair production and reliably well known in the Standard Model.  
 
Two basic ways to probe new physics can be investigated. The first one is a  
model independent analysis, in which the effects of new physics appear as new  
terms in an effective Lagrangian describing the interactions of the third  
family with gauge bosons and Higgs \cite{GounarisPR,WhisnantYYZ,YangY}. 
The effects due to the interactions between quarks and gauge  
bosons will be visible at LEP2, $e^+e^-$ next linear colliders and
the Tevatron whereas dimension 6 CP violating operators affect the
transverse polarisation asymmetry of the top quark. The second way is to
consider a new theory which contains the Standard Model at low energies. A
possible framework is supersymmetry. In the Minimal Supersymmetric Model
with $R$ parity conservation, the single top production at Tevatron is
enhanced by a few percent due to gluino, squarks, higgs, charginos and
neutralinos corrections, the magnitude being sensitive to $\tan \beta$
\cite{LI}. The decays $t \rightarrow cV$ with $V= g, Z,\gamma$, which are
small in magnitude in the Standard Model (BR $ \simeq 10^{-10}-10^{-12}$),
may be enhanced by a few orders of magnitude in the MSSM \cite{Couture}. If
the stop and the charged Higgs are light enough new top decays are possible
\cite{Hosch}. Our purpose is to investigate the effects of $R$ parity
violation. The superpotential contains three types of new terms: 
\be 
W_{\not \! R} = \lambda_{ijk} L_i L_j{\bar E}_k +  \lambda'_{ijk}  
L_i Q_j {\bar D}_k + \lambda^{''}_{ijk} {\bar U}_i {\bar D}_j {\bar D}_k   
\label{eq1} 
\ee 
the first two terms violating the leptonic number and the last the 
baryonic one. Here $L$ and $E$ are isodoublet and isosinglet lepton, 
$Q$ and $D$ are isodoublet and isosinglet quark super-fields, the 
indices $i$, $j$ and $k$ take values for the three lepton 
and quark families. In the following we shall assume that $R$-parity 
violation arises from one of these terms only. 
 
The feasibility of single top quark production via squark and slepton 
exchanges  
to probe several combinations of $R$ parity violating couplings at
hadron colliders has been studied \cite{DR,Datta,Oakes}. The LHC is
better at probing the $B$ violating couplings $\lambda^{\prime \prime}$
whereas the Tevatron and the LHC have a similar sensitivity to $\lambda'$
couplings. We perform a complete and detailed  
study including for the signal all channels using a Monte Carlo  
generator based on Pythia 6.1 \cite{sjostrand}, taking into account all the 
backgrounds and including the ATLAS detector  
response using ATLFAST 2.0 \cite{atlfast}.    
 
The paper is organised as follows. 
Section~\ref{theory} is devoted to an evaluation of  
the different subprocesses contributing to single top production (standard  
model, squark, slepton and charged Higgs exchanges). The potential of the LHC  
to discover or put limits on $R$-parity violating interactions is  
given in section~\ref{experiment}.    
 
\section{Subprocesses contributing to single top production} 
\label{theory}

The $R$-parity violating parts of the Lagrangian that contribute to single  
top production are:  
\be 
L_{\not {\! R}} =  \lambda'_{ijk} {\tilde e}^i_L {\bar d}^k_R u^j_L -    
\lambda^{''}_{ijk} ({\tilde d}^k_R {\bar u}^i_L d^j_L + {\tilde d}^j_R  
({\bar d}^k_L)^c  u^i_L) + h.c.  
\label{eq2} 
\ee 
The superscript $c$ corresponds to charge conjugation. 
There are altogether 27 and 9 $\lambda'_{ijk}$ and $\lambda''_{ijk}$ Yukawa 
couplings, respectively. 
The most suppressed couplings are $ \lambda'_{111}$,  
$ \lambda'_{133}$, $\lambda^{''}_{112}$,   
$\lambda^{''}_{113}$  (see \cite{gdr} for detailed up to date reviews of the
existing bounds). In order to fix
the kinematical variables, the reaction we consider is 
\be 
u_i(p_1) + d_j (p_2) \rightarrow t(p_3) + b(p_4) \;\; , 
\label{eq3} 
\ee 
the $p_k$ being the 4-momenta of the particles and the indices $i$ and 
$j$ refer to the generations of the $u$ and $d$-type quarks. 
 
We first discuss valence-valence (VV) or sea-sea (SS) subprocesses (this 
notation refers to the proton-proton collisions at the LHC, but the 
calculation is valid in general). The SM squared amplitude due to W  
exchange in   
$\hat u$-channel~\footnote{The ``hat'' symbol refers to the usual 
Mandelstam variables for the process at the parton level.}  
is suppressed by the Kobayashi-Maskawa matrix elements $V_{u_ib} V_{td_j}$:  
\be 
|M^{VV}_{WW}|^2 \; =\; g^4 \, |V_{u_ib}|^2\, |V_{td_j}|^2\,  
\frac{1}{(\hat u-m_W^2)^2 + m_W^2\Gamma_W^2}\;\; p_1\cdot p_2\; 
p_3\cdot p_4 ,
\label{prima} 
\ee 
where $g$, $m$ and $\Gamma$ denote the weak coupling constant, the mass 
and the width of the exchanged particle. 
The $H^{\pm}$ exchange in $\hat u$-channel is included in the calculation but 
numerically suppressed by the quark masses and the mixing matrix  
elements for the charged Higgs sector $K_{u_ib}K_{td_j}$(under  
the assumption $K = V$): 
\bea 
|M^{VV}_{{H^\pm} {H^\pm}}|^2 \; &=&\; \frac{g^4}{16\, m_W^4} \, |K_{u_ib}|^2\, 
|K_{td_j}|^2\,  
\frac{1}{(\hat u-m_{H^\pm}^2)^2 + m_{H^\pm}^2\Gamma_{H^\pm}^2}\nonumber\\ 
&& 
[(v^2_{bu_i}+a^2_{bu_i})\, p_1\cdot p_4\; +\;
(v^2_{bu_i}-a^2_{bu_i})\,m_bm_{u_i}]  [(v^2_{d_jt}+a^2_{d_jt})\, p_2 \cdot
p_3\; +\; (v^2_{d_jt}-a^2_{d_jt})\,m_{d_j}m_t] , 
\eea 
$v_{ud}$ and $a_{ud}$  are respectively the vector and axial vector couplings 
of $H^\pm$ to quarks:  
\be 
v_{ud} = m_d\tan \beta + m_u\cot \beta \;\;\;\;\;\; 
a_{ud} =m_d\tan \beta -m_u\cot \beta .
\ee
The interference term between the W and $H^\pm$ is:  
\bea 
2\, {\cal R}e(M^{VV}_{W {H^\pm}}) \; &=&\; -\, 
\frac{g^4}{8\, m_W^2} \, |V_{u_ib}|\;|V_{td_j}|\;|K_{u_ib}|\;|K_{td_j}|\,  
\frac{(\hat u-m_W^2)(\hat u-m_{H^\pm}^2) + 
m_W\Gamma_Wm_{H^\pm}\Gamma_{H^\pm} } 
{[(\hat u-m_W^2)^2 + m_W^2\Gamma_W^2] 
[(\hat u-m_{H^\pm}^2)^2 + m_{H^\pm}^2 
\Gamma_{H^\pm}^2] 
}\nonumber\\ 
&\times & [
(v_{bu_i}+a_{bu_i})(v_{d_jt}+a_{d_jt})\, m_{b} m_{d_j}\; p_1 \cdot p_3\;  +\;  
(v_{bu_i}+a_{bu_i})(v_{d_jt}-a_{d_jt})\, m_b m_t\; p_1 \cdot p_2\; 
\nonumber\\ 
&+& 
(v_{bu_i}-a_{bu_i})(v_{d_jt}+a_{d_jt})\, m_{u_i} m_{d_j}\; p_3 \cdot p_4\;
+\;   (v_{bu_i}-a_{bu_i})(v_{d_jt}-a_{d_jt})\, m_{u_i} m_t\; p_2 \cdot p_4 ] .
\eea 
The scalar slepton exchange in $\hat u$-channel is taken into account 
but appears to be suppressed within our assumptions on the $\lambda'$ 
couplings (see below): 
\be 
|M^{VV}_{{{\tilde e}^k_L}{{\tilde e}^k_L}}|^2 \; =\;   
\lambda_{ki3}^{\prime 2} \; \lambda_{k3j}^{\prime 2} 
\frac{1}{(\hat u-m_{{\tilde e}^k_L}^2)^2 + m_{{\tilde e}^k_L}^2 
\Gamma_{{\tilde e}^k_L}^2}\;\; p_1\cdot p_4\; p_2\cdot p_3 .
\label{eq6} 
\ee 
The interference term between scalar slepton and W reads:  
\bea 
2\, {\cal R}e(M^{VV}_{W{{\tilde e}^k_L}})  = -\, 
g^2 \, |V_{u_ib}|\;|V_{td_j}|\, \lambda_{ki3}'\; \lambda_{k3j}' 
\frac{(\hat u-m_W^2)(\hat u-m_{{\tilde e}^k_L}^2) + 
m_W\Gamma_Wm_{{\tilde e}^k_L}
\Gamma_{{\tilde e}^k_L} } 
{[(\hat u-m_W^2)^2 + m_W^2\Gamma_W^2] 
[(\hat u-m_{{\tilde e}^k_L}^2)^2 + m_{{\tilde e}^k_L}^2 
\Gamma_{{\tilde e}^k_L}^2] 
}\; m_{d_j} \, m_b\; p_1\cdot p_3 .
\eea 
The interference term between scalar slepton and $H^\pm$, which is suppressed, 
reads:  
\bea 
2\, {\cal R}e(M^{VV}_{{{\tilde e}^k_L}{H^\pm}})  &=&
\frac{g^2}{4\, m_W^2} \, \lambda_{ki3}' \; \lambda_{k3j}'\; |K_{u_ib}|\;  
|K_{td_j}|\,  
\frac{(\hat u-m_{H^\pm}^2)(\hat u-m_{{\tilde e}^k_L}^2) + m_{H^\pm}
\Gamma_{H^\pm}m_{{\tilde e}^k_L}\Gamma_{{\tilde e}^k_L} } 
{[(\hat u-m_{H^\pm}^2)^2 + m_{H^\pm}^2\Gamma_{H^\pm}^2] 
[(\hat u-m_{{\tilde e}^k_L}^2)^2 + m_{{\tilde e}^k_L}^2 
\Gamma_{{\tilde e}^k_L}^2] 
}\nonumber\\ 
&\times & 
[(v_{bu_i}+a_{bu_i})(v_{d_jt}+a_{d_jt})\; p_1\cdot p_4\; p_2\cdot p_3\;  +\;  
(v_{bu_i}+a_{bu_i})(v_{d_jt}-a_{d_jt})\, m_{d_j} m_t\; p_1\cdot p_4\;
\nonumber\\
&+&
(v_{bu_i}-a_{bu_i})(v_{d_jt}+a_{d_jt})\, m_{u_i}m_b\; p_2\cdot p_3\; +\;  
(v_{bu_i}-a_{bu_i})(v_{d_jt}-a_{d_jt})\, m_{u_i}m_{d_j} m_b m_t ] .
\eea 
The down type squark exchange in $\hat s$-channel squared amplitude is  
dominant and given by:  
\be 
|M^{VV}_{{{\tilde d}^k_R}{{\tilde d}^k_R}}|^2 \; =\;  \frac{4}{3}\;16\; 
\lambda_{ijk}^{'' 2}\; \lambda_{33k}^{'' 2} 
\frac{1}{(\hat s-m_{{\tilde d}^k_R}^2)^2 + m_{{\tilde d}^k_R}^2 
\Gamma_{{\tilde d}^k_R}^2}\;\; p_1\cdot p_2\; p_3\cdot p_4 .
\label{eq7} 
\ee 
The corresponding interference terms are:  
\be 
2\, {\cal R}e(M^{VV}_{W{{\tilde d}^k_R}})  = -\frac{2}{3}\; 8\, 
g^2 \, |V_{u_ib}|\; |V_{td_j}|\, \lambda_{ijk}^{''}\; \lambda_{33k}^{''} 
\frac{(\hat u-m_W^2)(\hat s-m_{{\tilde d}^k_R}^2) 
+ m_W\Gamma_Wm_{{\tilde d}^k_R}
\Gamma_{{\tilde d}^k_R} } 
{[(\hat u-m_W^2)^2 + m_W^2\Gamma_W^2] 
[(\hat s-m_{{\tilde d}^k_R}^2)^2 + m_{{\tilde d}^k_R}^2 
\Gamma_{{\tilde d}^k_R}^2] }\; p_1\cdot p_2 \; p_3\cdot p_4 ,
\ee
and:  
\bea 
2\, {\cal R}e(M^{VV}_{{{\tilde d}^k_R}{H^\pm}})  &=& \frac{2}{3}\, 
\frac{g^2}{2\, m_W^2} \, \lambda_{ijk}^{''}\; \lambda_{33k}^{''}\; 
|K_{u_ib}|\; |K_{td_j}|\,  
\frac{(\hat u-m_{H^\pm}^2)(\hat s-m_{{\tilde d}^k_R}^2) + m_{H^\pm}
\Gamma_{H^\pm}m_{{\tilde d}^k_R}\Gamma_{{\tilde d}^k_R} } 
{[(\hat u-m_{H^\pm}^2)^2 + m_{H^\pm}^2\Gamma_{H^\pm}^2] 
[(\hat s-m_{{\tilde d}^k_R}^2)^2 + m_{{\tilde d}^k_R}^2 
\Gamma_{{\tilde d}^k_R}^2]} 
\nonumber\\ 
&\times & 
[(v_{bu_i}+a_{bu_i})(v_{d_jt}+a_{d_jt})\, m_{b}m_{d_j}\, p_1\cdot p_3\; +\;  
(v_{bu_i}+a_{bu_i})(v_{d_jt}-a_{d_jt})\, m_bm_t\; p_1\cdot p_2 \nonumber\\ 
&+& 
(v_{bu_i}-a_{bu_i})(v_{d_jt}+a_{d_jt})\, m_{u_i}m_{d_j}\; p_3\cdot p_4\; +\;  
(v_{bu_i}-a_{bu_i})(v_{d_jt}-a_{d_jt})\, m_{u_i}m_t\; p_2\cdot p_4 ] .
\eea 
 
Let us now take into account the subprocesses involving valence-sea (VS) 
quarks. The SM squared amplitude due to W exchange in the $\hat s$-channel, 
being proportional to ${(V_{u_id_j} V_{tb})}^2$ is dominant for quarks of 
the same generation. It reads: 
\be 
|M^{VS}_{WW}|^2 \; =\; g^4 \, |V_{u_id_j}|^2\, |V_{tb}|^2\,  
\frac{1}{(\hat s-m_W^2)^2 + m_W^2\Gamma_W^2}\;\; p_1\cdot p_4\; 
p_2\cdot p_3 .
\label{eq8} 
\ee 
The charged Higgs contribution in the $\hat s$-channel is suppressed by  
the quark masses of the initial state. The squared amplitude is:  
\bea 
|M^{VS}_{{H^\pm}{H^\pm}}|^2 \; &=&\; \frac{g^4}{16\, m_W^4} \, 
|K_{u_id_j}|^2\, |K_{tb}|^2\,  
\frac{1}{(\hat s-m_{H^\pm}^2)^2 + m_{H^\pm}^2\Gamma_{H^\pm}^2}\nonumber\\ 
&\times & 
[(v^2_{d_ju_i}+a^2_{d_ju_i})\, p_1\cdot p_2\; -\; (v^2_{d_ju_i}-a^2_{d_ju_i})\,
m_{d_j}m_{u_i}] [(v^2_{bt}+a^2_{bt})\; p_3\cdot p_4\; -\;
(v^2_{bt}-a^2_{bt})\,m_bm_t] . 
\eea 
The interference term between W and $H^\pm$ is:  
\bea  
2\, {\cal R}e(M^{VS}_{W{H^\pm}})  &=& -\, 
\frac{g^4}{8\, m_W^2} \, |V_{u_id_j}|\; |V_{tb}|\;|K_{u_id_j}|\;|K_{tb}|\,  
\frac{(\hat s-m_W^2)(\hat s-m_{H^\pm}^2) + m_W\Gamma_Wm_{H^\pm}\Gamma_{H^\pm}} 
{[(\hat s-m_W^2)^2 + m_W^2\Gamma_W^2] 
[(\hat s-m_{H^\pm}^2)^2 + m_{H^\pm}^2 
\Gamma_{H^\pm}^2] }\nonumber \\ 
&\times &
[(v_{d_ju_i}+a_{d_ju_i})(v_{bt}+a_{bt})\, m_bm_{d_j}\; p_1\cdot p_3\; -\;  
(v_{d_ju_i}+a_{d_ju_i})(v_{bt}-a_{bt})\, m_{d_j}m_t\; p_1\cdot p_4 \nonumber\\
&-&
(v_{d_ju_i}-a_{d_ju_i})(v_{bt}+a_{bt})\, m_{b}m_{u_i}\; p_2\cdot p_3\; +\;  
(v_{d_ju_i}-a_{d_ju_i})(v_{bt}-a_{bt})\, m_{u_i}m_t\; p_2\cdot p_4\; ] .
\eea 
 
Concerning $R$ parity violating terms, slepton exchange in $\hat s$-channel 
and down type squark exchange in the $\hat u$-channel contribute:  
\bea 
|M^{VS}_{{{\tilde e}^k_L}\; {{\tilde e}^k_L}}|^2 &=&   
\lambda_{kij}^{\prime 2}.\lambda_{k33}^{\prime 2} 
\frac{1}{(\hat s-m_{{\tilde e}^k_L}^2)^2 + m_{{\tilde e}^k_L}^2 
\Gamma_{{\tilde e}^k_L}^2}\;\; p_1\cdot p_2\; p_3\cdot p_4 
\nonumber \\ 
|M^{VS}_{{{\tilde d}^k_R}{{\tilde d}^k_R}}|^2 &=& \frac{4}{3}\; 16\; 
\lambda_{i3k}^{'' 2}\; \lambda_{3jk}^{'' 2} 
\frac{1}{(\hat u-m_{{\tilde d}^k_R}^2)^2 + m_{{\tilde d}^k_R}^2 
\Gamma_{{\tilde d}^k_R}^2}\;\; p_1\cdot p_4\; p_2\cdot p_3 .
\eea 
The interference terms involving the scalar lepton are:  
\be
2\, {\cal R}e(M^{VS}_{W{{\tilde e}^k_L}}) \; =\; -\, 
g^2 \, |V_{u_id_j}| \;|V_{tb}|\, \lambda_{kij}'\; \lambda_{k33}' 
\frac{(\hat s-m_W^2)(\hat s-m_{{\tilde e}^k_L}^2) 
+ m_W\Gamma_Wm_{{\tilde e}^k_L}
\Gamma_{{\tilde e}^k_L} } 
{[(\hat s-m_W^2)^2 + m_W^2\Gamma_W^2] 
[(\hat s-m_{{\tilde e}^k_L}^2)^2 + m_{{\tilde e}^k_L}^2 
\Gamma_{{\tilde e}^k_L}^2] 
}\; m_d \; m_b\; p_1\cdot p_3 ,
\ee 
and:  
\bea 
2\, {\cal R}e(M^{VS}_{{{\tilde e}^k_L} {H^\pm}})  &=& 
\frac{g^2}{4\, m_W^2} \, \lambda_{kij}'\; \lambda_{k33}'\;  
|K_{u_id_j}|\; |K_{tb}|\,  
\frac{(\hat s-m_{H^\pm}^2)(\hat s-m_{{\tilde e}^k_L}^2) + m_{H^\pm}
\Gamma_{H^\pm}m_{{\tilde e}^k_L}\Gamma_{{\tilde e}^k_L} } 
{[(\hat s-m_{H^\pm}^2)^2 + m_{H^\pm}^2\Gamma_{H^\pm}^2] 
[(\hat s-m_{{\tilde e}^k_L}^2)^2 + m_{{\tilde e}^k_L}^2 
\Gamma_{{\tilde e}^k_L}^2] 
}\nonumber\\ 
&\times & 
[(v_{d_ju_i}+a_{d_ju_i})(v_{bt}+a_{bt})\; p_1\cdot p_2\; p_3\cdot p_4\; -\;  
(v_{d_ju_i}+a_{d_ju_i})(v_{bt}-a_{bt})\; m_tm_b\; p_1.p_2 \nonumber \\ 
&-&
(v_{d_ju_i}-a_{d_ju_i})(v_{bt}+a_{bt})\, m_{d_j}m_{u_i}\; p_3 \cdot p_4\; +\;  
(v_{d_ju_i}-a_{d_ju_i})(v_{bt}-a_{bt})\, m_{u_i}m_{d_j}m_bm_t ] .
\eea 
The interference terms involving the scalar quark are:  
\be  
2\, {\cal R}e(M^{VS}_{W {{\tilde d}^k_R}}) \; =\; -\frac{2}{3}\; 8\, 
g^2 \, |V_{u_id_j}|\; |V_{tb}|\, \lambda_{i3k}^{''}\; \lambda_{3jk}^{''}  
\frac{(\hat s-m_W^2)(\hat u-m_{{\tilde d}^k_R}^2) + m_W
\Gamma_Wm_{{\tilde d}^k_R}\Gamma_{{\tilde d}^k_R} } 
{[(\hat s-m_W^2)^2 + m_W^2\Gamma_W^2] 
[(\hat u-m_{{\tilde d}^k_R}^2)^2 + m_{{\tilde d}^k_R}^2 
\Gamma_{{\tilde d}^k_R}^2] 
}\; p_1\cdot p_4\; p_2\cdot p_3 
\ee 
and:  
\bea  
2\, {\cal R}e(M^{VS}_{{{\tilde d}^k_R} {H^\pm}}) \; &=& \frac{2}{3}\, 
\frac{g^2}{2\, m_W^2} \, \lambda_{i3k}^{''}\; \lambda_{3jk}^{''}\; 
|K_{u_id_j}|\; |K_{tb}|\,  
\frac{(\hat s-m_{H^\pm}^2)(\hat u-m_{{\tilde d}^k_R}^2) 
+ m_{H^\pm}\Gamma_{H^\pm}m_{{\tilde d}^k_R}\Gamma_{{\tilde d}^k_R} } 
{[(\hat s-m_{H^\pm}^2)^2 + m_{H^\pm}^2\Gamma_{H^\pm}^2] 
[(\hat u-m_{{\tilde d}^k_R}^2)^2 + m_{{\tilde d}^k_R}^2 
\Gamma_{{\tilde d}^k_R}^2] } \nonumber \\
&\times & 
[(v_{d_ju_i}+a_{d_ju_i})(v_{bt}+a_{bt})\, m_bm_{d_j}\; p_1\cdot p_3\; -\;  
(v_{d_ju_i}+a_{d_ju_i})(v_{bt}-a_{bt})\, m_{d_j}m_t\; p_1\cdot p_4 \nonumber \\
&-&
(v_{d_ju_i}-a_{d_ju_i})(v_{bt}+a_{bt})\, m_{b}m_{u_i}\; p_2.p_3\; +\;  
(v_{d_ju_i}-a_{d_ju_i})(v_{bt}-a_{bt})\, m_{u_i}m_t\; p_2.p_4\; ] .
\label{ulti} 
\eea 
The dominant terms are the squared amplitude due to $\tilde e$  
exchange, and for initial quarks of the same generation ($i=j$), 
the interference between $W$ and $\tilde d$. The result is sensitive 
to the interference term only if 
the product of $\lambda''$ couplings is large (around $10^{-1}$). 
For subprocesses involving quarks of different generations in the 
initial state the situation is more complex and all amplitudes have to 
be taken into account.  
 
The resonant $\hat s$-channel processes have been studied in 
\cite{Oakes}, for first family up and down quarks. For the $B$-violating 
couplings, the study of $\hat s$-channels $cd \rightarrow$  $\tilde s$ and 
$cs\rightarrow \tilde d$ can also be found in \cite{Oakes}. 
The $\hat u$-diagram 
has been studied at the Tevatron for the first family of up and down 
quarks \cite{Datta}. 
 
In the present note we have improved previous calculations for LHC  
because we have included all contributions to single top production.  
Since the dominant terms are those considered in the literature,  
our complete evaluation validates the approximations done in previous papers. 
 
\section{Detection of single top production through $R$-parity violation 
at the LHC} 
\label{experiment}
 
We have carried out the feasibility study to detect  
single top production through $R$-parity violation at the LHC by measuring the  
{$l\nu bb$ final state using the following procedure. 
 
First, we have implemented the partonic $2 \to 2$ 
cross sections calculated using Eqs. (\ref{prima})--(\ref{ulti})  
in the PYTHIA event generator. Providing PYTHIA with 
the flavour and the momenta of the initial partons 
using a given parton distribution function (p.d.f.)\footnote{ 
We have used the CTEQ3L p.d.f. We checked that the use of different sets of 
parton distribution functions induces an uncertainty in the results which is
around the percent level. This in no way affects our conclusions.} 
it then generates complete final states including initial and final state 
radiations and hadronization.  
 
The generated events were implemented in ATLFAST to simulate the response of
the  ATLAS detector. In particular, 
isolated electrons, photons were smeared with the detector 
resolution in the pseudo-rapidity 
range of $|\eta|<2.5$. 
In the same way and the same $\eta$ region the measured parameters 
of the isolated and non-isolated  muons were simulated. Finally, 
a simple fixed cone algorithm (of radius $R=0.4$) was used to 
reconstruct the parton jets. 
The minimum transverse energy of a jet was set at 15 GeV. 
According to the expected b-tagging performance of the ATLAS detector
\cite{TDR} for low luminosity at the LHC we have assumed a $60\%$ b-tag 
efficiency for a factor 100 of rejection against light jets.  
 
The same procedure was applied to the SM background with the 
exception that we used besides PYTHIA also the ONETOP~\cite{onetop}  
event generator. 
 
The integrated luminosity for one year at low luminosity at the LHC is 
taken to be 10 fb$^{-1}$. 
 
The number of signal events depends on the mass and the width of the 
exchanged sparticle, and on the value of the Yukawa couplings 
(see Section~\ref{theory}). 
We assume that only one type of Yukawa coupling is nonzero, i.e. either 
sleptons ($\lambda '\ne 0$) or squarks ($\lambda ''\ne 0$) are exchanged. 
The width of the the exchanged sparticle is a sum of the widths 
due to $R$-parity conserving and $R$-parity violating decays: 
\begin{equation} 
\Gamma_{tot} = \Gamma_{R} + \Gamma_{\RPV} 
\label{eq:totwidth} 
\end{equation} 
where $\Gamma_{\RPV}$ is given by  
\begin{equation} 
\Gamma_{\RPV}(\tilde q^i_R\longrightarrow q^jq^k)=
\frac{(\lambda''_{ijk})^2}{2\pi} 
\frac{(M^2_{\tilde q^i_R}-M^2_{top})^2}{M^3_{\tilde q^i_R}} 
\label{eq:sq-width} 
\end{equation} 
for the squarks, and it is given by 
\begin{equation} 
\Gamma_{\RPV}(\tilde l^i_L\longrightarrow q^j\bar q^k)=
\frac{3(\lambda'_{ijk})^2}{16\pi} 
\frac{(M^2_{\tilde l^i_L}-M^2_{top})^2}{M^3_{\tilde l^i_L}}
\label{eq:sl-width} 
\end{equation} 
for the sleptons. The number of signal events depends also on the 
flavour of the initial partons through their p.d.f. In 
Table~\ref{tb:sq_xsec} we display the 
total cross section values for different initial parton flavours in the 
case of exchanged squarks of mass of 600 GeV and of $R$-parity 
conserving width 
$\Gamma_{R}$ = 0.5 GeV. We took for all $\lambda '' = 10^{-1}$, which yields a 
natural width of the squark which is smaller than the experimental
resolution. Table~\ref{tb:sl_xsec} contains the same information for
slepton  exchange ( $\lambda ' = 10^{-1}$, for a slepton of mass of 250 GeV
and a width of $\Gamma_{R}$ = 0.5 GeV).  Other processes are not quoted
because the small value of the limits of their couplings prevents their
detection.     

In order to study the dependence of the signal on the mass and the width 
of the exchanged particle we have fixed the couplings to $10^{-1}$ and have 
chosen three different masses for the 
exchanged squarks: 300, 600 and 900 GeV, respectively. For each mass value 
we have chosen two different $\Gamma_R$: 0.5 and 20 GeV, respectively. 
For the first case $\Gamma_{\RPV}$ dominates, whereas in the last 
one, when $\Gamma_{tot} \approx \Gamma_R$, the single 
top-production cross section decreases by a factor $\sim 10$.  
We have considered here the $ub$ parton initial state, since this has 
the highest cross section value.  
Besides, we have also generated events with a $cd$ initial partonic state 
and an exchanged $\tilde s$-quark of mass of 300 GeV, for comparison 
with the simulation presented in Ref.~\cite{Oakes}. 
 
In order to study the dependence on the parton initial state we have fixed 
the mass of the exchanged squark to 600 GeV and its width with 
$\Gamma_R$ = 0.5 GeV and varied the initial state according to the 
first line of Table~\ref{tb:sq_xsec}. 
 
Finally, for the exchanged sleptons we have studied only one case, 
namely the $u\bar d$ initial state with a mass and width of the exchanged 
slepton of 250 GeV and 0.5 GeV, respectively. 
In each case we have generated about $10^5$ signal events. 
 
The different types of background considered are listed in 
Table~\ref{tb:sigmabg} 
together with their estimated cross sections. 
The irreducible backgrounds are single top production through a  
virtual $W$ (noted $W^*$), or through $W$-gluon fusion.  
$W$-gluon fusion is the dominant process (for a detailed study 
see \cite{stelzer}). 
A $Wbb$ final state can be obtained either in direct production 
or through $Wt$ or $t \bar t$ production.   
Finally, the reducible background consists of $W$-n$j$ events where two of the 
jets are misidentified as $b$-jets. 
 
We have used the ONETOP\cite{onetop} event generator to simulate 
the $W$-gluon fusion process. For the other backgrounds we have 
used PYTHIA. We have generated from one thousand ($W^*$) to several million 
events ($t\bar t$) depending on the importance of the background. 
   
The separation of the signal from the background is based on the 
presence of a resonant structure of the $tb$ final state 
in the case of the signal. The background does not show such a 
structure as it is illustrated in Fig~1.  
 
In the process to reconstruct the $tb$ final state first we 
reconstruct the top quark. 
The top quark can be reconstructed from the $W$ and from one of the  
$b$-quarks in the final state, requiring that their invariant mass satisfy 
\begin{eqnarray*} 
150 \leq M_{Wb} \leq 200 & \mbox{GeV}. 
\end{eqnarray*} 
The $W$ is in turn reconstructed from either of the two decay channels: 
\begin{eqnarray*} 
W &\rightarrow& u \bar d \nonumber \\ 
W &\rightarrow& l \nu . 
\end{eqnarray*} 
Here we have considered only the latter case which gives a better  
signature due to the presence of a high $p_t$ lepton 
and missing energy. The former case suffers from multi-jets event  
backgrounds.  
As we have only one neutrino, its longitudinal momentum  can be reconstructed 
by using the $W$ and top mass constraints. 
The procedure used is the following : \\ 
- we keep events with two b-jets of $p_t \ge 40$ GeV, with one lepton of 
$p_t \ge 25$ GeV, with $E_t^{miss} \ge  35$ GeV  and with a jet multiplicity 
$\le 3$,\\ 
- we reconstruct the longitudinal component ($p_z$) of the neutrino by  
requiring $M_{l\nu}$ = $M_W$. 
This leads to an equation with twofold ambiguity on $p_z$.  \\ 
- More than 80$\%$ of the events have at 
least one solution for $p_z$. In case of two solutions, we calculate $M_{l\nu  
b}$ for each of the two b-jets and we keep the $p_z$ that minimises  
$|M_{top}-M_{l\nu b}|$.\\ 
- we keep only events where $150  \le   M_{l\nu b} \le 200$ GeV.\\ 

Next, the reconstructed top quark is combined with the 
$b$ quark not taking part 
in the top reconstruction. An example of the invariant mass distribution of 
the $tb$ final state is shown in Fig.~2.
 
In order to reduce the the $t\bar t$ background to a manageable level, 
we need to apply a strong jet veto on the third jet by requiring that its 
$p_t$ should be $\le 20$ GeV. 

The invariant mass distribution of the $tb$ final state after having applied 
this cut is shown in Fig.~3. The signal to background
ratio is clearly increased in comparison to Fig.~2. 

Once an indication for a signal is found, we count the number of 
signal ($N_s$) 
and background ($N_b$) events in an interval corresponding to  
2 standard deviations around the signal peak for an integrated luminosity of 
30 fb$^{-1}$. Then we rescale the signal 
peak by a factor $\alpha$ such that  
\[ 
             N_s/\sqrt{N_b} = 5. 
\] 
By definition the scale-factor $\alpha$ determines the limit of  
sensitivity for the lowest value of the  
$\lambda''$ ($\lambda'$) coupling we can test with the LHC: 
\[ 
          \lambda''_{ijk}\cdot\lambda''_{lmn} \le 0.01\cdot \sqrt\alpha. 
\]       
 
In Table~\ref{tb:lambda_ub} we show the limits obtained for the 
combinations of $\lambda''_{132}\lambda''_{332}$ for different masses 
and widths of the exchanged $\tilde s$-quark. Also shown are the current 
limits assuming a mass for $\tilde m_f$ = 100 GeV, 
the number of signal and background 
events, as well as the experimentally observable 
widths of the peak ($\Gamma_{exp}$). In Fig.~4 
we compare our results with those obtained in Ref.~\cite{Oakes} 
for $m_{\tilde s} = 300$ GeV, and a $cd$ initial state, 
using parton-level simulation. 
We ascribe the lower efficiency of this analysis to the more detailed and 
realistic detector simulation employed.  

In Table~\ref{tb:lambda_600} we compile the sensitivity limit of the 
bilinear combination of the different Yukawa couplings one can 
obtain after 3 years of LHC run with low luminosity, if the exchanged 
squark has a mass of 600 GeV. For its width we consider 
$\Gamma_R$ = 0.5 GeV and a component $\Gamma_{\RPV}$ given 
by Eq.(\ref{eq:sq-width}).  
 
For the exchanged sleptons (cf Table~\ref{tb:sl_xsec}) 
we have calculated the sensitivity limit of the 
bilinear combination of the different Yukawa couplings 
only for the most favourable case, i.e. for the $u\bar d$} 
partonic initial state. 
We obtain 4.63$\times10^{-3}$ for the 
limits on $\lambda'_{11k}\lambda'_{k33}$ (in comparison with the limit of 
2.8$\times10^{-3}$ obtained by Oakes {\it et al.}). 
For those cases where the exchanged squark (slepton) might be 
discovered at the LHC we have made an estimate on the precision 
one can determine 
its mass. For this purpose, we have subtracted the background under the mass 
peak and fitted a Gaussian curve on the remaining signal. 
This procedure is illustrated in Fig.~5 for the 
case of 600 GeV squark mass and $ub$ partonic initial state. 
For the assumed value of the coupling constant, the error on 
the mass determination 
is dominated by the $1\%$ systematic uncertainty on the jet energy scale 
in ATLAS\cite{TDR}. 
 
\section{Conclusions} 
We have studied single top production through $R$-parity violating 
Yukawa type couplings, at the LHC. 
 
We have considered all $2\rightarrow 2$ partonic processes at tree-level, 
including interference terms. The calculated $2\rightarrow 2$ partonic 
cross sections have been implemented in PYTHIA to generate complete 
particle final states. A fast particle level simulation was used to 
obtain the response of the ATLAS detector. We have taken into account 
all important SM backgrounds. 
 
We have studied the signal-to-background ratio as a function of the 
initial partonic states, the exchanged sparticle mass and width, 
and of the value of the Yukawa couplings. 
 
At the chosen value of the coupling constants ($\sim 10^{-1}$), 
significant signal-to-background ratio was obtained only in the 
$\hat s$-channel, in the $tb$ $(l\nu bb)$ invariant mass distribution, 
around the mass of the exchanged sparticle, if \\ 
\indent $(i)$ the exchanged sparticle is a squark, and \\ 
\indent $(ii)$ its width due to $R$-parity conserving decay is of the 
order of a GeV. \\ 
In this case we obtain a significance of $S = N_s/\sqrt{N_b} > 5$ 
for the whole mass range investigated (300 -- 900 GeV) for an 
integrated luminosity of 30 fb$^{-1}$. This means, 
that squarks ($\tilde d$ or $\tilde s$) with narrow width might be 
discovered at the LHC. The experimental mass resolution would allow 
to measure the squark mass with a precision of $\sim 1$\%. 
 
Conversely, if no single top production above the SM expectation 
is observed at the LHC, after 3 years of running at low luminosity, 
the experimental limit on the quadratic combination of the $\lambda ''$ 
couplings can be lowered by at least one order of magnitude, for narrow width 
squarks. In the case of slepton exchange significant signal-to-background 
ratio can be obtained for $u\bar d$ partonic initial state, 
i.e. for the combination of the $\lambda_{11k}'\lambda_{k33}'$ couplings. 
Due to the lower rate, as compared to squark exchange, in the absence of 
a signal, the current limit can be improved only by a factor of two. 
The difference between the significance in our study and  
the one in Ref.~\cite{Oakes} 
can be explained by the different degree of detail in the 
simulation process.     

\section*{Acknowledgements.} 
We thank S. Ambrosanio and S. Lola for useful comments on the manuscript.
A.D. acknowledges the support of a ``Marie Curie'' TMR research 
fellowship of the European Commission under contract ERBFMBICT960965 in 
the first stage of this work. J.-M. V. thanks the ``Alexander von 
Humboldt Foundation'' for financial support.

\section*{Figures}

\begin{figure}
\epsfxsize=9cm
\centerline{\epsffile{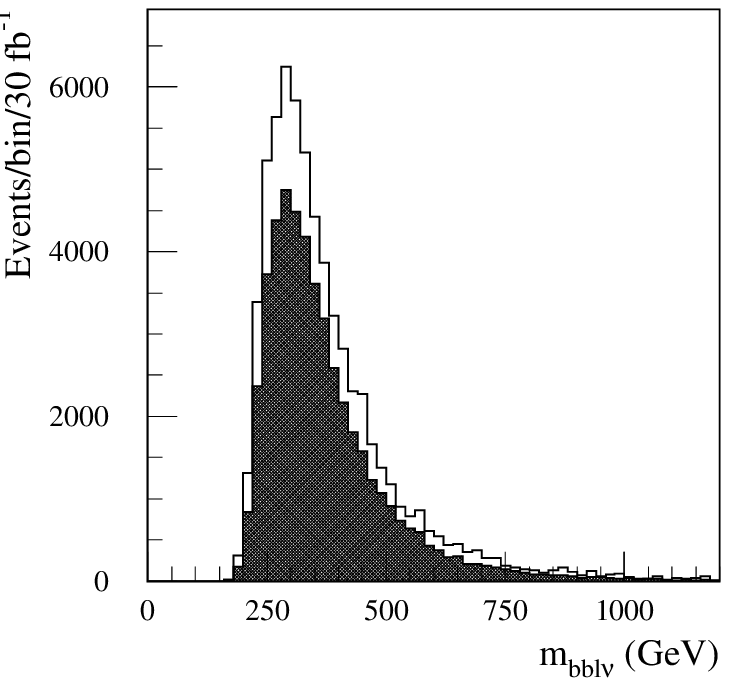}}
\noindent
{\bf Fig. 1} - {Invariant mass distribution of $l\nu bb$ for the  
backgrounds after three years at LHC at low luminosity. The $t\bar t$ 
background dominates (dashed histogram).} 
\end{figure}

\begin{figure}
\epsfxsize=9cm
\centerline{\epsffile{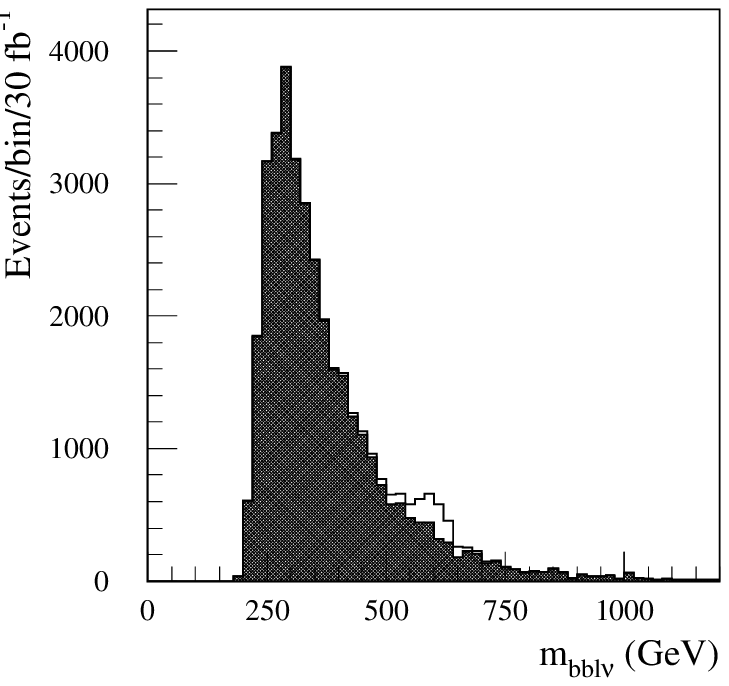}}
\noindent
{\bf Fig. 2} - {Invariant mass distribution of $l\nu bb$ combination 
for the signal and backgrounds (dashed histogram) after three years of 
LHC run at low luminosity.  
The signal corresponds to an exchanged $\tilde s$-quark of 600 GeV 
mass and 0.5 GeV width. The initial partons are $ub$ and the 
$\lambda''$ couplings are set to $10^{-1}$.}
\end{figure}

\begin{figure}
\epsfxsize=9cm
\centerline{\epsffile{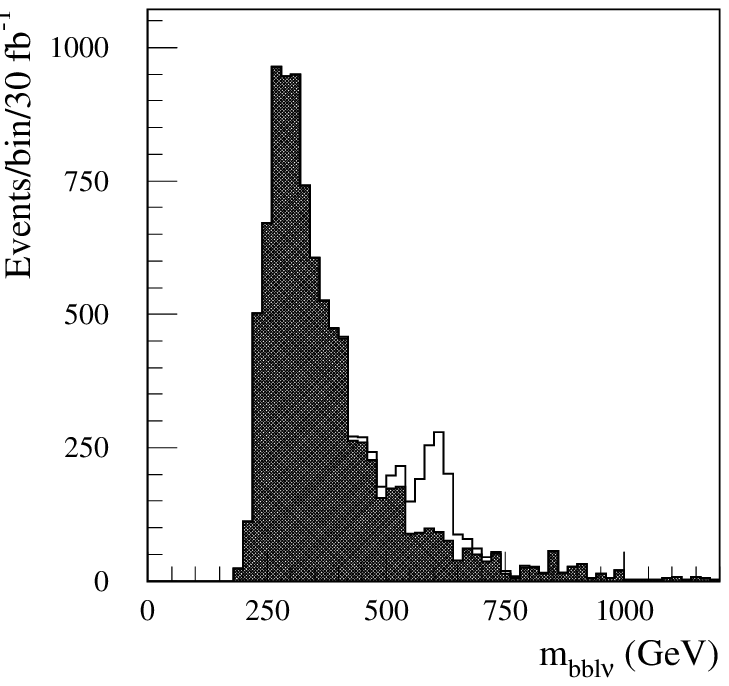}}
\noindent
{\bf Fig. 3} - {Invariant mass distribution of $l\nu bb$ for the signal and 
backgrounds (dashed histogram) after three years at LHC at low luminosity 
after having applied the cuts.}
\end{figure}

\begin{figure}
\epsfxsize=9cm
\centerline{\epsffile{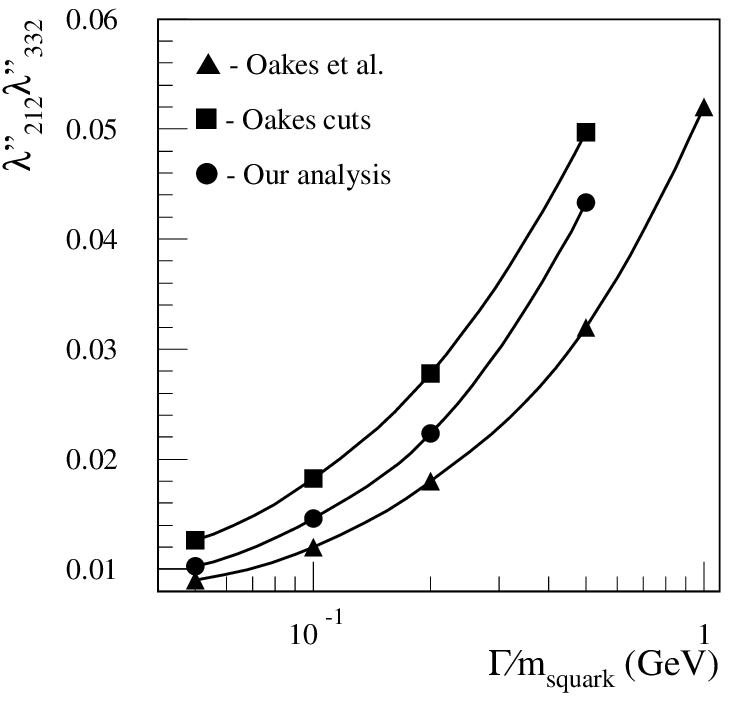}}
\noindent
{\bf Fig. 4} - {Sensitivity limits for the values of the 
$\lambda''_{212}\lambda''_{332}$ 
Yukawa couplings we obtain for the $cd$ initial state at the LHC 
after 1 year with low luminosity, for an exchanged $\tilde s$-quark 
of mass of 300 GeV (circles). The result obtained by 
Oakes {\it et al.}, is also shown (triangles). The squares indicate a result 
obtained by applying the cuts used by Oakes {\it et al.} on our sample.} 
\end{figure}

\begin{figure}
\epsfxsize=9cm
\centerline{\epsffile{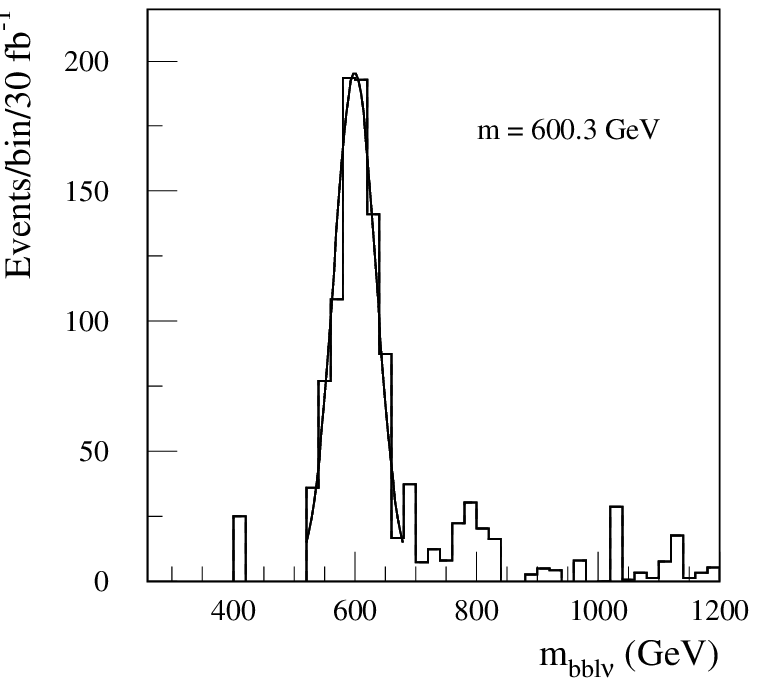}}
\noindent
{\bf Fig. 5} - {The background subtracted mass distribution fitted with 
a Gaussian in case of an exchanged $\tilde s$-quark of 600 GeV for a 
$ub$ initial parton state. It corresponds to 3 years of LHC run with 
low luminosity. } 
\end{figure}

\newpage
\section*{Tables}

\mediumtext
\begin{table} 
\centering 
\begin{tabular}{|c|c|c|c|c|c|}
Initial partons   & $cd$    &  $cs$    &  $ub$   &  \multicolumn{2}{ c|}{$cb$} \\ 
\tableline 
Exchanged particle   
&$\tilde s$ & $\tilde d$ &$ \tilde s$ & $ \tilde d$ & $ \tilde s$ 
\\ \tableline 
Couplings    
& $\lambda''_{212}\lambda''_{332}$   & $\lambda''_{212}\lambda''_{331}$   
& $\lambda''_{132}\lambda''_{332}$   & $\lambda''_{231}\lambda''_{331}$   
& $\lambda''_{232}\lambda''_{332}$   \\ \tableline 
Cross section in pb    
& 3.98  & 1.45   & 5.01  &  \multicolumn{2}{ c|} {0.659}
\end{tabular} 
\caption{Total cross-section in pb for squark exchange in the $\hat s$-channel 
for a squark of mass of 600 GeV assuming $\Gamma_{R}$ = 0.5 GeV.} 
\label{tb:sq_xsec} 
\end{table} 

\begin{table} 
\centering 
\begin{tabular}{|c|c|c|c|c|c|c|}
Initial partons & 
$u\bar d$ & $u\bar s$ & $c\bar d$ & $c\bar s$ & $u\bar b$ & $c\bar b$ \\
\tableline
Couplings
& $\lambda'_{11k}\lambda'_{k33}$   & $\lambda'_{12k}\lambda'_{k33}$   
& $\lambda'_{21k}\lambda'_{k33}$   & $\lambda'_{22k}\lambda'_{k33}$   
& $\lambda'_{13k}\lambda'_{k33}$   & $\lambda'_{23k}\lambda'_{k33}$   
\\ \tableline 
Cross section in pb    
& 7.05  & 4.45   & 2.31  &  1.07  & 2.64  & 0.525
\end{tabular} 
\caption{Total cross-section in pb for slepton exchange in the $\hat s$-channel 
for a slepton of mass of 250 GeV assuming $\Gamma_{R}$ = 0.5 GeV.} 
\label{tb:sl_xsec} 
\end{table} 

\narrowtext
\begin{table} 
\centering 
\begin{tabular}{cc}
Background  &  $\sigma\times BR$ (pb) \\ \tableline
$W^*$ & 2.2\\
gluon fusion & 54\\
$Wt$ & 17\\
$t\bar t$ & 246\\ 
$Wbb$ & 66.6\\ 
$Wjj$ & 440
\end{tabular} 
\caption{$\sigma \times$ Branching Ratio for backgrounds.} 
\label{tb:sigmabg} 
\end{table} 

\mediumtext
\begin{table} 
\centering 
\begin{tabular}{|c|c|c|c|c|c|c|}
{$m_{\tilde s}$ (GeV) } & \multicolumn{2} {c|} {300}&\multicolumn{2} {c|} 
{600}&\multicolumn{2} {c|} {900} \\ \tableline 
$\Gamma_R$ (GeV)  & 0.5 & 20 & 0.5 & 20 & 0.5 & 20 \\ \tableline 
$N_s$            & 6300 & 250 & 703 & 69 & 161 & 22 \\ \tableline 
$N_b$            & 4920 & 5640& 558 & 1056& 222& 215 \\ \tableline 
$\Gamma_{exp}$ (GeV)    & 24.3 & 30.5 & 37.5 & 55.6 & 55.4 & 62.1 \\ \tableline 
 Limits on $\lambda''\times\lambda''$ & 2.36$\times10^{-3}$&1.21$\times10^{-2}$ 
& 4.10$\times10^{-3}$& 1.51 
$\times10^{-2}$& 6.09$\times10^{-3}$& 2.09$\times10^{-2}$
\end{tabular} 
\caption{Limits for the values of the $\lambda''_{132}\lambda''_{332}$ 
Yukawa couplings for an integrated luminosity of 30 fb$^{-1}$. For the 
other quantities see the text. Current limit is 6.25$\times10^{-1}$.}  
\label{tb:lambda_ub} 
\end{table}  

\begin{table} 
\centering 
\begin{tabular}{|c|c|c|c|c|c|}
Initial partons   & $cd$    &  $cs$    &  $ub$   &  \multicolumn{2}{ c|}{$cb$} \\ 
\tableline 
Exchanged particle   &$\tilde s$ & $\tilde d$ &$ \tilde s$ & $ \tilde d$ & $ \tilde s$ 
\\ \tableline 
Couplings
& $\lambda''_{212}\lambda''_{332}$   & $\lambda''_{212}\lambda''_{331}$   
& $\lambda''_{132}\lambda''_{332}$   & $\lambda''_{231}\lambda''_{331}$   
& $\lambda''_{232}\lambda''_{332}$   \\ \tableline 
$N_s$   & 660  & 236 & 703 & \multicolumn{2}{ c|} {96}  \\ \tableline 
$N_b$   & \multicolumn{5}{ c|} {558}  \\ \tableline 
$\Gamma_{exp}$ (GeV) & 38.5  & 31.3 & 37.5 &   \multicolumn{2}{ c|} {40.1} \\ \tableline 
Limits on $\lambda''\times\lambda''$   & 4.26$\times10^{-3}$ & 7.08$\times10^{-3}$  
& 4.1$\times10^{-3}$ &  
\multicolumn{2}{ c|} {1.11$\times10^{-2}$}
\end{tabular} 
\caption{Limits on the Yukawa couplings for an exchanged squark of mass 
600 GeV assuming $\Gamma_R$ = 0.5 GeV, for an integrated luminosity 
of 30 fb$^{-1}$. Current limit is 6.25$\times10^{-1}$.} 
\label{tb:lambda_600} 
\end{table}

\end{document}